# Exploiting Reconfigurable Intelligent Surfaces in Edge Caching: Joint Hybrid Beamforming and Content Placement Optimization

Yingyang Chen, *Member, IEEE*, Miaowen Wen, *Senior Member, IEEE*, Ertugrul Basar, *Senior Member, IEEE*, Yik-Chung Wu, *Senior Member, IEEE*, Li Wang, *Senior Member, IEEE*, and Weiping Liu, *Member, IEEE*

## Abstract

Edge caching can effectively reduce backhaul burden at core network and increase quality-of-service at wireless edge nodes. However, the beneficial role of edge caching cannot be fully realized when the offloading link is in deep fade. Fortunately, the impairments induced by wireless propagation environments could be renovated by a reconfigurable intelligent surface (RIS). In this paper, a new RIS-aided edge caching system is proposed, where a network cost minimization problem is formulated to optimize content placement at cache units, active beamforming at base station and passive phase shifting at RIS. After decoupling the content placement subproblem with hybrid beamforming design, we propose an alternating optimization algorithm to tackle the active beamforming and passive phase shifting. For active beamforming, we transform the problem into a semidefinite programming (SDP) and prove that the optimal solution of SDP is always rank-1. For passive phase shifting, we introduce block coordinate descent method to alternately optimize the auxiliary variables and the RIS phase shifts. Further, a conjugate gradient algorithm based on manifold optimization is proposed to deal with the

Y. Chen and W. Liu are with the Department of Electronic Engineering, College of Information Science and Technology, Jinan University, Guangzhou 510632, China (e-mail: chenyy@jnu.edu.cn; wpl@jnu.edu.cn).

M. Wen is with the School of Electronic and Information Engineering, South China University of Technology, Guangzhou 510640, China (e-mail: eemwwen@scut.edu.cn).

E. Basar is with the Communications Research and Innovation Laboratory (CoreLab), Department of Electrical and Electronics Engineering, Koc University, 34450 Istanbul, Turkey (e-mail: ebasar@ku.edu.tr).

Y.-C. Wu is with the Department of Electrical and Electronic Engineering, The University of Hong Kong, Hong Kong (e-mail: ycwu@eee.hku.hk).

L. Wang is with the School of Electronic Engineering, Beijing University of Posts and Telecommunications (BUPT), Beijing 100876, China (email: liwang@bupt.edu.cn).

non-convex unit-modulus constraints. Numerical results show that our RIS-aided edge caching design can effectively decrease the network cost in terms of backhaul capacity and power consumption.

# I. INTRODUCTION

## A. Motivation and Scope

Nowadays, the driving forces of the exponential growth in mobile network traffic have been fundamentally shifted from the steady increase in demand for conventional *connection-centric* communications to the explosion of *content-centric* ones [1]. Considering the characteristics of cache-able content as well as skewed content popularity, there is consensus today that caching can increase network performance, reduce expenditures for operators, and improve quality-of-service for users [2], [3].

Exploiting the concept of caching to support content delivery over wireless networks is referred to as **edge caching**, i.e., caching at a base station (BS) or mobile devices [4]. Intrinsically, edge caching trades scarce wireless communication bandwidth and transmission power with storage resources by exploiting the time-reversal characteristic of the traffics into the system. Caching also enables edge computing capabilities by pre-installing necessary computing datasets at the wireless edge nodes [5]. However, edge caching is fundamentally different from caching in a content delivery network (CDN), since it disperses content files at the wireless edge. Predominantly, the content transmission link over the wireless medium is far from perfect, which leads to uncertain caching performance. For example, the mobile devices located at the cell edge typically suffer from a low delivery rate, and their received content is prone to a low successful probability. On the other hand, the signals from multiple BSs can act as interference, but also can be exploited through cooperative multiple point (CoMP) transmission [6], multicast beamforming [7], and interference alignment (IA) [8] to improve the delivery quality and efficiency. Therefore, the edge caching designs require joint consideration of wireless transmissions and caching strategies.

With the theoretical and experimental breakthrough in micro-electro-mechanical systems and meta-materials, communications with reconfigurable intelligent surfaces (RISs), also called in-telligent reflecting surfaces (IRSs), have recently been proposed as a powerful solution to enhancing the spectral efficiency (SE) and energy efficiency (EE) of wireless networks [9]–[11]. In particular, an RIS comprises an array of low-cost reflecting elements to proactively configure the end-to-end wireless propagation channel. Compared to legacy relaying systems, the RIS shapes the incoming signal by controlling the phase shift of each reflecting element instead of employing a power amplifier, and is capable of supporting full-duplex and full-band transmissions inherently. Furthermore, as electromagnetic materials, RISs are easy to be coated on existing structures such as building facades, vehicle windows, and indoor walls, which largely reduces the complexity of deployment. By shaping wireless propagation environments proactively, RIS-empowered communication could be a key enabler for improving caching performance achieved at wireless edge nodes.

## B. Related Works

*1) Edge Caching in Wireless Networks:* The design of wireless transmission techniques changes significantly in the presence of edge caching. Xu *et al.* [12] showed that caching at both transmitters and receivers can turn an interference channel to a cooperative X-multicast channel. In [8], the authors used zero-forcing (ZF) and IA techniques in the delivery phase to exploit the presence of identical file portions at multiple transmitters and hence reap multiple benefits from caching. In [7], the authors introduced a content-centric multicast beamforming design for content delivery in a cache-enabled radio access network. Poularakis *et al.* [13] optimized caching policies based on multicast transmissions to reduce energy costs. The caching performance analysis and optimization in stochastic wireless networks have also attracted great attention [6], [14]–[16], where the node locations are generally modeled as independent spatial random processes, e.g., Poisson Point Process, and advanced stochastic geometry tools can be resorted to [17]. Wen *et al.* [6] introduced CoMP into small BS (SBS) caching in a downlink large-scale heterogeneous network (HetNet). In [14], the authors exploited full-duplex relaying to boost wireless caching in a two-tier HetNet, where the success probability was derived in closed-form. Liu *et al.* [15] investigated the optimal caching policy to maximize the success probability and area SE in a cache-enabled HetNet. In [16], the authors investigated probabilistic content placement to control cache-based channel selection diversity and network interference. There are some works devoted to exploring the interplay between edge caching and other emerging wireless technologies and services, e.g., full duplex [18]–[20], non-orthogonal transmission [21], [22], unmanned aerial vehicles (UAVs) [23], and vehicular communications [24]. Against this background, there is a paucity of literature for enhancing the caching performance by invoking RISs.

*2) RIS Empowered Wireless Networks:* To exploit the gains provided by RISs, there is a number of works applying signal enhancement with the reflection path [25]–[30]. Hybrid beam-forming consisting of active transmit/receive beamforming at transceiver and passive phase shifting at RIS was often developed to fully reap the multiple-antenna gain, where one of the difficulties lies in the non-convex unit-modulus constraints induced by the phase shifts of RISs [9]. In [25], Wu *et al.* minimized the transmit power for an RIS-aided multiple-input single-output (MISO) system by jointly optimizing the transmit beamforming and the reflect pattern, where the passive beamforming was designed by invoking the semidefinite relaxation (SDR) approach. Ning *et al.* [26] maximized the SE of an RIS-assisted multiple-input multiple-output (MIMO) system. The passive beamforming was solved by utilizing alternating direction method of multipliers (ADMM) after taking the amplitude of each reflection coefficient into consideration. Huang *et al.* [27] maximized the sum rate in an RIS-aided MISO downlink communication, where the non-convexity in the RIS matrix was tackled with the aid of a majorization-minimization (MM) method. Di *et al.* [28] investigated hybrid beamforming in a downlink RIS multi-user system where the discrete feature of phase shifts were considered. Yu *et al.* [29] proposed to reformulate the RIS phase shift deign problem into an equivalent rank-constrained problem, which was further solved by a difference of convex (DC) method. More recently, low complexity RIS phase adjustment algorithms were also explored in [30].

Besides, there is another group of researches to align the reflected signals of RISs for signal cancellation at certain terminals, which are often applied in physical layer security (PLS) and interference cancellation [31]–[33]. For example, Lyu *et al.* [31] investigated an RIS jamming scenario, where RISs act as jammers for attack a legitimate communication without using internal energy. In [32], the authors invoked RISs at the cell boundary to assist the downlink transmission to cell-edge users whilst mitigating the inter-cell interference. Hou *et al.* [33] designed a passive beamforming weight in RIS-aided MIMO non-orthogonal multiple access (NOMA) networks, where the inter-cluster interference can be eliminated.

There are also some contributions devoted to integrating RISs with diverse wireless networks and other emerging technologies, such NOMA networks [33]–[35], mobile edge computing (MEC) [36], simultaneous wireless information and power transfer (SWIPT) [37], [38], UAV wireless networks [39], vehicular communications [40], and machine learning [41]. All above impressive contributions motivate us to exploit the benefits of RISs in edge caching systems.

*C. Contributions and Organizations*

Our main contributions in this paper are detailed as follows.

- Firstly, we develop a new RIS-aided edge caching design for the first time in literature, and formulate a network cost minimization problem. Specifically, we first propose an RIS-aided edge caching system to assist content offloading for mobile users. The network cost minimization problem is formulated to optimize hybrid beamforming and content placement, where the network cost consists of both the backhaul capacity and the transmission power cost. The hybrid beamforming considered consists of active beamforming at BS and passive phase shifting at RIS. Owing to the non-convex unit-modulus constraints and the coupling of multiple optimization variables, the network-cost-minimization problem cannot be solved in a straightforward manner. By analyzing the problem structure, we decouple the content placement subproblem with the hybrid beamforming design.

- Secondly, we propose an alternating optimization algorithm to decouple active beamforming at BS with passive beamforming at RIS. When fixing the passive beamformer to optimize the active one, we transform the problem into a semidefinite programming (SDP) by applying SDR. We prove that the optimal solution of the SDP is the solution of the primal active beamforming problem exactly. When fixing the active beamformer to optimize the passive one, we confront with a feasibility problem with non-convex unit-modulus constraints. By introducing auxiliary variables based on a penalty function method, we transform the signal-to-interference-noise-ratio (SINR) constraints into the objective function of the feasibility problem. A block coordinate descent method is exploited to alternately optimize the auxiliary variables and the RIS phase shifts.

- Thirdly, we introduce an effective conjugate gradient algorithm based on manifold optimization to deal with the non-convex unit-modulus constraints. We show that the unit-modulus constraints of all phase shifters constitute a complex circle manifold, i.e., a Riemannian manifold. Then, we can transform to optimize the passive phase shifts in a Riemannian manifold, where the original unit-modulus constraints can be easily guaranteed. The optimized results can be obtained through the predefined retraction and mapping operations between the Riemmannian manifold and the Euclidean space. Goldstein criterion line search and Flecther-Reeves equation are employed to guarantee converging to a locally optimal solution.

- Finally, we present numerical validations and evaluations. Numerical results show that our RIS-aided edge caching design can effectively decrease the network cost in terms of backhaul capacity and power consumption. Specifically, the power cost can be decreased by installing more passive reflecting elements on RIS. Meanwhile, our proposed content placement design achieves lower backhaul cost than other existing caching strategies. Furthermore, simulation results show that the RIS location should be carefully chosen, and the path loss exponent of RIS-related links has prominent impact on the achievable performance. It is shown that the RIS had better to be deployed in an open scenario so as to reap more performance gain.

The rest of this paper is organized as follows. In Section II, we establish the system model and formulate the network cost minimization problem. An alternating optimization method is developed in Section III. In Section IV, we propose a block coordinate descent method for passive beamforming. Our numerical results are discussed in Section V. Finally, the conclusion of this article is presented in Section VI.

*Notation*: Uppercase and lowercase bold-faced letters indicate matrices and vectors, respectively. Calligraphic letters denote sets. Let $(\cdot)^T$, $(\cdot)^*$ and $(\cdot)^H$ refer to the transpose, conjugate and conjugate-transpose operations, respectively. The $l_2$-norm of a vector $\mathbf{a}$ is shown by $\|\mathbf{a}\|_2$, and $(\mathbf{a})_i$ represents the $i$-th element of $\mathbf{a}$. Notation $\mathbf{A} \succeq \mathbf{0}$ represents that $\mathbf{A}$ is a positive semidefinite matrix. The ring of complex numbers is denoted as $\mathbb{C}$, whilst $\mathbf{A} \in \mathbb{C}^{M \times N}$ indicates that $\mathbf{A}$ is a complex-element matrix with dimensions $M \times N$. An $N \times N$ dimensional identity matrix is denoted as $\mathbf{I}_N$, and $\mathrm{diag}(\cdot)$ denotes the diagonalization operation. The range space and null space of matrix $\mathbf{A}$ are denoted as $\mathcal{R}(\mathbf{A})$ and $\mathcal{N}(\mathbf{A})$, respectively. The real part of a complex argument is noted by $\mathrm{Re}\{\cdot\}$, and $\circ$ denote the Hadamard (element-wise) product between two matrices. Finally, $x \sim \mathcal{CN}(\mu, \sigma^2)$ indicates that the random variable $x$ obeys a complex Gaussian distribution with mean $\mu$ and variance $\sigma^2$.

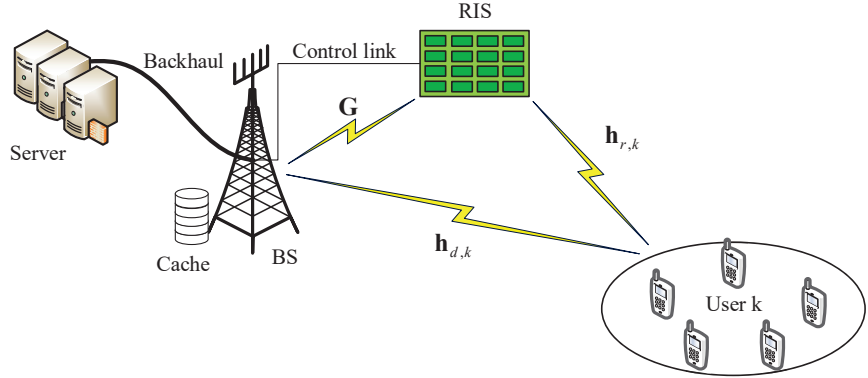

Fig. 1.   RIS-aided edge caching system.

## II. System Model and Problem Formulation

### A. System Model

As illustrated in Fig. 1, we consider the downlink transmission of a cache-enabled radio access network with an RIS. There exist an $M$ multiple-antenna BS, $K$ single-antenna mobile users, and an RIS equipped with $N$ passive reflecting elements. In practice, the RIS is configured by the BS through a control link. The BS has local cache with a limited storage size. Meanwhile, the BS is also connected to the server via a high-capacity backhaul link and can access a database that contains a total number of $F$ files with equal size. Let $\mathcal{K} = \{1, \cdots, K\}$ denote the set of mobile users, and $\mathcal{N} = \{1, \cdots, N\}$ represent the set of RIS elements. The channels between BS and RIS, the user-$k$ and BS, the user-$k$ and RIS are respectively noted as $\mathbf{G} \in \mathbb{C}^{N \times M}$, $\mathbf{h}_{d,k} \in \mathbb{C}^{M \times 1}$, and $\mathbf{h}_{r,k} \in \mathbb{C}^{N \times 1}$.

### B. Cache Model

Let $\mathcal{F} = \{1, \cdots, F\}$ denote the set of file indices in the database, where each file is assumed to have normalized size of $1$, and its popularity obeys the Zipf distribution with skewness factor $\varepsilon$. The local storage size of the BS is denoted as $S_0$ ($S_0 < F$), which represents the maximum number of files it can cache. We assume that the BS applies a probabilistic caching strategy, that is, the BS caches a content randomly with deterministic probability. More specifically, we define a content placement vector $\mathbf{c} = \{c_1, \cdots, c_f, \cdots, c_F\}$, where $c_f \in [0, 1]$ indicates the probability that the $f$-th content file is cached in the BS. Due to the limited cache size, $\sum_{f \in \mathcal{F}} c_f \leq S_0$ should be satisfied. For the users, $b_f^k$ denotes the probability that user $k$ requests file $f$. For simplicity, we assume that the request distributions for different users are uniform, i.e., $b_f^1 = b_f^2 = \cdots = b_f^K \triangleq b_f$. Further, we assume that users request content files according to their popularity, and the request probability follows a Zipf distribution with skewness factor $\varepsilon$, i.e., $b_f = \frac{f^{-\varepsilon}}{\sum_{i=1}^{F} i^{-\varepsilon}}$. In general, a large value of $\varepsilon$ means more user requests are concentrated on fewer popular files.

During the period of content fetching, if content $f$ has been cached in BS's local storage, users can access the content directly without costing the backhaul link. Otherwise, it needs to fetch this content from BS via backhaul. Since the data rate of fetching a content needs to be as large as the content-delivery target rate $R_k^0$, we model the total backhaul cost as $\sum_{f=1}^{F} \sum_{k=1}^{K} (1 - c_f) b_f R_k^0$.

### C. Communication Model

Let $\mathbf{p}_k \in \mathbb{C}^{M \times 1}$ denote the precoding vector at BS for user $k$, and $\mathbf{P} = [\mathbf{p}_1, \cdots, \mathbf{p}_k, \cdots, \mathbf{p}_K] \in \mathbb{C}^{M \times K}$ denote the active precoding matrix at BS. Further, let $\mathbf{\Theta} = \text{diag} \left\{ e^{j\theta_1}, \cdots, e^{j\theta_n}, \cdots, e^{j\theta_N} \right\}$ denote the reflection coefficient matrix at the RIS, where $\theta_n \in [0, 2\pi)$ is the phase shift of the $n$-th reflecting element. The signal received at user $k$ from both of the BS-user and BS-RIS-user channels can be expressed as

$$y_k = \left( \mathbf{h}_{d,k}^H + \mathbf{h}_{r,k}^H \mathbf{\Theta} \mathbf{G} \right) \mathbf{Ps} + z_k \tag{1}$$

$$= \sum_{l=1}^{K} \left( \mathbf{h}_{d,k}^H + \mathbf{h}_{r,k}^H \mathbf{\Theta} \mathbf{G} \right) \mathbf{p}_l s_l + z_k \tag{2}$$

where $z_k \sim \mathcal{CN} \left( 0, \sigma_0^2 \right)$ is the i.i.d Gaussian noise at the receiver of user $k$, and $\sigma_0^2$ is the noise power spectral density. Therefore, the SINR of user $k$ is given by

$$\gamma_k = \frac{\left| \left( \mathbf{h}_{d,k}^H + \mathbf{h}_{r,k}^H \mathbf{\Theta} \mathbf{G} \right) \mathbf{p}_k \right|^2}{\sum\limits_{l=1, l \neq k}^{K} \left| \left( \mathbf{h}_{d,k}^H + \mathbf{h}_{r,k}^H \mathbf{\Theta} \mathbf{G} \right) \mathbf{p}_l \right|^2 + \sigma_0^2 B} \tag{3}$$

where $B$ is the system bandwidth.

### D. Problem Formulation

In this paper, we aim to optimize active beamformer $\mathbf{P}$, passive beamformer $\mathbf{\Theta}$ and content placement vector $\mathbf{c}$ to minimize the total network cost, which consists of both the backhaul capacity and the transmission power. To this end, we formulate the optimization problem as

$$\mathcal{P}_0: \min_{\{c_f\}, \{\mathbf{p}_k\}, \{\theta_n\}} \sum_{f=1}^{F} \sum_{k=1}^{K} (1 - c_f) \, b_f R_k^0 + \eta \sum_{k=1}^{K} \|\mathbf{p}_k\|_2^2 \tag{4a}$$

$$s.t. \quad \gamma_k \geq \gamma_k^0, \; \forall k \in \mathcal{K}, \tag{4b}$$

$$c_f \in [0, 1], \; \forall f \in \mathcal{F} \tag{4c}$$

$$\sum_{f=1}^{F} c_f \leq S_0 \tag{4d}$$

$$0 \leq \theta_n < 2\pi, \; \forall n \in \mathcal{N} \tag{4e}$$

where $\gamma_k^0 = 2^{R_k^0/B} - 1$ is the SINR requirement with respect to the content-delivery target rate of user $k$, and $\eta > 0$ is the pricing factor to trade power for backhaul capacity.

Constraint (4b) is a non-convex minimum required SINR constraint, while constraint (4d) considers the local storage limit at BS, and constraint (4e) defines the interval of phase shifts, which is a non-convex unit-modulus constraint. Hence the formulated problem is non-convex and cannot be solved in a straightforward manner.

## III. Alternating Optimization Development

In this section, we propose an alternating optimization method to deal with the formulated problem $\mathcal{P}_0$. The key idea is to decouple the content placement optimization with the beamforming design, and then decouple the active beamforming with the passive phase shifts as so to alternately update each other till convergence.

### A. Content Placement Optimization

For the content placement part in the alternating optimization, both of the passive and active beamformers are fixed. Hence $\mathcal{P}_0$ is transformed into

$$\begin{aligned} \mathcal{P}_1: \min_{\{c_f\}} \quad & \sum_{f=1}^{F} (1 - c_f) \, b_f \\ s.t. \quad & c_f \in [0, 1], \; \forall f \in \mathcal{F} \\ & \sum_{f=1}^{F} c_f \leq S_0 \end{aligned} \tag{5}$$

Above program is convex with respect to the content placement vector $\mathbf{c} = \{c_1, \cdots, c_f, \cdots, c_F\}$, and can be solved by KKT conditions easily.

## B. Beamforming Optimization

When fixing the content placement to optimize beamforming, $\mathcal{P}_0$ is transformed into

$$
\begin{aligned}
\mathcal{P}_2 : \min_{\{\mathbf{p}_k\},\{\theta_n\}} & \ \sum_{k=1}^{K} \|\mathbf{p}_k\|_2^2 \\
s.t. \quad & \gamma_k \geq \gamma_k^0, \ \forall k \in \mathcal{K} \\
& 0 \leq \theta_n \leq 2\pi, \ \forall n \in \mathcal{N}.
\end{aligned} \tag{6}
$$

In this sequel, we also turn to an alternating methodology to deal with $\mathcal{P}_2$.

*1) Active beamforming:* When fixing the passive beamformer $\boldsymbol{\Theta}$ to optimize the active beamformer $\mathbf{P}$, $\mathcal{P}_2$ becomes

$$
\begin{aligned}
\mathcal{P}_{2-\mathrm{I}} : \min_{\{\mathbf{p}_k\}} & \ \sum_{k=1}^{K} \|\mathbf{p}_k\|_2^2 \\
s.t. \quad & \gamma_k \geq \gamma_k^0, \ \forall k \in \mathcal{K}.
\end{aligned} \tag{7}
$$

This problem in the present form is nonconvex since the inequality constraint function apparently is nonconvex. Next, let us reformulate it into an SDP.

By denoting $\mathbf{f}_k^H = \mathbf{h}_{d,k}^H + \mathbf{h}_{r,k}^H \boldsymbol{\Theta} \mathbf{G}$, the inequality constraints in (7) can be re-expressed as

$$
\frac{1}{\gamma_k^0} \left| \mathbf{f}_k^H \mathbf{p}_k \right|^2 \geq \sum_{l=1, l \neq k}^{K} \left| \mathbf{f}_k^H \mathbf{p}_l \right|^2 + \sigma_0^2, \, k = 1, \cdots, K. \tag{8}
$$

To cope with $\mathcal{P}_{2-\mathrm{I}}$, we reformulate the complex-variable problem into an SDP by applying SDR. Specifically, we replace the rank-one matrix $\mathbf{p}_k \mathbf{p}_k^H$ by a general-rank positive semidefinite (PSD) matrix $\mathbf{P}_k$ of $M \times M$ dimensions. Then the following SDP problem is yielded

$$
\begin{aligned}
\mathcal{P}_{2-\mathrm{I}}^{SDR} : \min_{\{\mathbf{P}_k\}} & \ \sum_{k=1}^{K} \mathrm{Tr}\left(\mathbf{P}_k\right) \\
s.t. \quad & \frac{1}{\gamma_k^0} \mathbf{f}_k^H \mathbf{P}_k \mathbf{f}_k \geq \sum_{l=1, l \neq k}^{K} \mathbf{f}_k^H \mathbf{P}_l \mathbf{f}_k + \sigma_0^2, \ \forall k \in \mathcal{K} \\
& \mathbf{P}_k \succeq \mathbf{0}, \ \forall k \in \mathcal{K}.
\end{aligned} \tag{9}
$$

***Lemma* 1:** The obtained optimal $\left\{\mathbf{P}_k^{(*)}\right\}$ of problem (9) is of rank one, i.e., $\mathbf{P}_k^{(*)} = \mathbf{p}_k^{(*)} \left(\mathbf{p}_k^{(*)}\right)^H$ ; thus $\left\{\mathbf{p}_k^{(*)}\right\}$ must be an optimal solution of problem (7).

*Proof:* The Lagrangian of problem (9) is shown as

$$
\begin{aligned}
\mathcal{L}\left(\{\mathbf{P}_k, \lambda_k, \mathbf{Z}_k\}\right) &= \sum_{k=1}^{K} \mathrm{Tr}\left(\left[\mathbf{I} - \frac{\lambda_k}{\gamma_k^0}\mathbf{f}_k\mathbf{f}_k^H - \mathbf{Z}_k\right]\mathbf{P}_k\right) + \sum_{k=1}^{K}\sum_{\substack{l=1\\l\neq k}}^{K}\mathrm{Tr}\left(\lambda_k\mathbf{f}_k\mathbf{f}_k^H\mathbf{P}_l\right) + \sum_{k=1}^{K}\lambda_k\sigma_0^2 \\
&= \sum_{k=1}^{K}\mathrm{Tr}\left(\left[\mathbf{I} - \frac{\lambda_k}{\gamma_k^0}\mathbf{f}_k\mathbf{f}_k^H - \mathbf{Z}_k\right]\mathbf{P}_k\right) + \sum_{l=1}^{K}\sum_{\substack{k=1\\k\neq l}}^{K}\mathrm{Tr}\left(\lambda_l\mathbf{f}_l\mathbf{f}_l^H\mathbf{P}_k\right) + \sum_{k=1}^{K}\lambda_k\sigma_0^2 \\
&= \sum_{k=1}^{K}\mathrm{Tr}\left(\left[\mathbf{I} - \frac{\lambda_k}{\gamma_k^0}\mathbf{f}_k\mathbf{f}_k^H + \sum_{\substack{l=1\\l\neq k}}^{K}\lambda_l\mathbf{f}_l\mathbf{f}_l^H - \mathbf{Z}_k\right]\mathbf{P}_k\right) + \sum_{k=1}^{K}\lambda_k\sigma_0^2
\end{aligned}
\tag{10}
$$

where $\lambda_k \in \mathbb{R}$ and $\mathbf{Z}_k \in \mathbb{H}^M$ are the dual variables associated with the constraints in (9). Let

$$
\mathbf{A}_k = \mathbf{I} - \frac{\lambda_k}{\gamma_k^0}\mathbf{f}_k\mathbf{f}_k^H + \sum_{\substack{l=1\\l\neq k}}^{K}\lambda_l\mathbf{f}_l\mathbf{f}_l^H - \mathbf{Z}_k
\tag{11}
$$

The dual function of problem (9) can be easily seen to be

$$
\begin{aligned}
&g\left(\{\lambda_k, \mathbf{Z}_k\}\right) \\
&= \inf_{\mathbf{P}_k \in \mathbb{C}^M}\mathcal{L}\left(\{\mathbf{P}_k, \lambda_k, \mathbf{Z}_k\}\right) \\
&= \inf_{\mathbf{P}_k \in \mathbb{C}^M}\sum_{k=1}^{K}\frac{1}{2}\left[\mathrm{Tr}\left(\mathbf{A}_k\mathbf{P}_k\right) + \mathrm{Tr}\left(\mathbf{A}_k^*\mathbf{P}_k^*\right)\right] + \sum_{k=1}^{K}\lambda_k\sigma_0^2 \\
&= \begin{cases} \sum_{k=1}^{K}\lambda_k\sigma_0^2, & \mathbf{A}_k = \mathbf{0}, \ \forall k \\ -\infty, & \text{otherwise.} \end{cases}
\end{aligned}
\tag{12}
$$

Thus, the dual of problem (9) is given by

$$
\begin{aligned}
&\max_{\{\lambda_k, \mathbf{Z}_k\}} \ \sum_{k=1}^{K}\lambda_k\sigma_0^2 \\
&s.t. \ \ \mathbf{A}_k = \mathbf{0}, \ \mathbf{Z}_k \succeq \mathbf{0}, \ \lambda_k \geq 0, \ \forall k.
\end{aligned}
\tag{13}
$$

Since problem (9) is convex and Slater's condition holds true, KKT conditions are the sufficient and necessary optimality conditions. Some KKT conditions needed in the proof are as follows

$$
\frac{1}{\gamma_k^0}\mathbf{f}_k^H\mathbf{P}_k^{(*)}\mathbf{f}_k \geq \sum_{l=1, l\neq k}^{K}\mathbf{f}_k^H\mathbf{P}_l^{(*)}\mathbf{f}_k + \sigma_0^2, \ k = 1, \cdots, K
\tag{14}
$$

$$
\mathbf{Z}_k^{(*)} = \mathbf{I} - \frac{\lambda_k^{(*)}}{\gamma_k^0}\mathbf{f}_k\mathbf{f}_k^H + \sum_{l=1, l\neq k}^{K}\lambda_l^{(*)}\mathbf{f}_l\mathbf{f}_l^H, \ k = 1, \cdots, K
\tag{15}
$$

$$
\mathbf{Z}_k^{(*)}\mathbf{P}_k^{(*)} = \mathbf{0}, \ k = 1, \cdots, K.
\tag{16}
$$

Note that $\lambda_k \geq 0$, $\mathbf{Z}_k^{(*)} \succeq \mathbf{0}$ and $\mathbf{P}_k^{(*)} \succeq \mathbf{0}$ are also KKT conditions.

Since $\mathbf{P}_k^{(*)} \neq \mathbf{0}$ (by (14)), the rank of $\mathbf{Z}_k^{(*)}$ must be less than or equal to $M-1$ to ensure that at least one dimension of null space exists (by (16)), i.e.,

$$\text{rank}\left(\mathbf{Z}_k^{(*)}\right) \leq M - 1. \tag{17}$$

By defining

$$\mathbf{B} = \mathbf{I} + \sum_{l=1, l \neq k}^{K} \lambda_l \mathbf{f}_l \mathbf{f}_l^H = \left(\mathbf{B}^{1/2}\right)^2 \succ \mathbf{0} \tag{18}$$

where $\mathbf{B}^{1/2} = \left(\mathbf{B}^{1/2}\right)^H \succ \mathbf{0}$, the rank of $\mathbf{Z}_k^{(*)}$ can be further inferred as follows

$$
\begin{aligned}
&\text{rank}\left(\mathbf{Z}_k^{(*)}\right) \\
&= \text{rank}\left(\mathbf{B} - \frac{\lambda_k^{(*)}}{\gamma_k^0} \mathbf{f}_k \mathbf{f}_k^H\right) \\
&= \text{rank}\left(\mathbf{B}^{1/2}\left[\mathbf{I} - \frac{\lambda_k^{(*)}}{\gamma_k^0} \mathbf{B}^{-1/2}\mathbf{f}_k \mathbf{f}_k^H \mathbf{B}^{-1/2}\right]\mathbf{B}^{1/2}\right) \\
&= \text{rank}\left(\mathbf{I} - \frac{\lambda_k^{(*)}}{\gamma_k^0} \mathbf{B}^{-1/2}\mathbf{f}_k \mathbf{f}_k^H \mathbf{B}^{-1/2}\right) \\
&\geq M - 1.
\end{aligned}
\tag{19}
$$

From (17) and (19), it can be inferred that $\text{rank}\left(\mathbf{Z}_k^{(*)}\right) = M - 1$. Then, by (16), we have

$$
\begin{aligned}
&\text{rank}\left(\mathbf{P}_k^{(*)}\right) \leq \dim\left(\mathcal{N}\left(\mathbf{Z}_k^{(*)}\right)\right) = M - \text{rank}\left(\mathbf{Z}_k^{(*)}\right) = 1 \\
&\Rightarrow \text{rank}\left(\mathbf{P}_k^{(*)}\right) = 1 \quad (\text{since } \mathbf{P}_k^{(*)} \neq \mathbf{0}).
\end{aligned}
$$

Therefore, the optimal solution $\left\{\mathbf{P}_k^{(*)}\right\}$ of the SDR problem (9) must yield the optimal solution $\left\{\mathbf{p}_k^{(*)}\right\}$ of the active beamforming problem (7) via the rank-one decomposition $\mathbf{P}_k^{(*)} = \mathbf{p}_k^{(*)}\left(\mathbf{p}_k^{(*)}\right)^H$. ∎

Till now, by solving (9) efficiently via off-the-self convex solvers and then applying Lemma 1, we can obtain optimized active beamformers $\left\{\mathbf{p}_k^{(*)}\right\}$ of program $\mathcal{P}_{2-I}$ in (7).

*2) Passive beamforming:* When fixing the active beamformer $\mathbf{P}$ to optimize the passive beamformer $\mathbf{\Theta}$, $\mathcal{P}_2$ reduces to

$$
\begin{aligned}
\mathcal{P}_{2-\text{II}} : \min_{\{\theta_n\}} \quad & \sum_{k=1}^{K} \|\mathbf{p}_k\|_2^2 \\
s.t. \quad & \gamma_k \geq \gamma_k^0, \ \forall k \in \mathcal{K} \\
& 0 \leq \theta_n \leq 2\pi, \ \forall n \in \mathcal{N}.
\end{aligned}
\tag{20}
$$

Apparently, $\mathcal{P}_{2-\text{II}}$ is a feasibility problem, where the objective function is independent of the variables [47]. It is clear that the difficulty of the problem solving mainly lies in the non-convex unit modulus constraints induced by the phase shifts. In the following section, we resort to a block coordinate descent algorithm to solve the feasibility problem above.

## IV. Block Coordinate Descent Method for Passive Beamforming

In the following, we propose a block coordinate descent algorithm to deal with passive beamforming problem above. Based on a penalty function method, we transform the SINR constraints into the objective function of the feasibility problem by introducing auxiliary variables. The block coordinate descent method is proposed to alternately optimize the auxiliary variables and the RIS phase shifts.

### A. Problem Reformulation

First, we introduce new auxiliary variables to transform the SINR constraints in (20) into the objective function. To this end, let

$$\left( \mathbf{h}_{d,k}^{H} + \mathbf{h}_{r,k}^{H} \mathbf{\Theta} \mathbf{G} \right) \mathbf{p}_l = x_{k,l} \tag{21}$$

and

$$\left( \mathbf{h}_{d,k}^{H} + \mathbf{h}_{r,k}^{H} \mathbf{\Theta} \mathbf{G} \right) \mathbf{p}_k = x_{k,k}. \tag{22}$$

Then the SINR constraints in (20) can be expressed as

$$\mathcal{P}_{2-\text{II}} : \quad \min_{\{\theta_n\}, \{x_{k,j}\}} \quad \sum_{k \in \mathcal{K}} \|\mathbf{p}_k\|_2^2 \tag{23a}$$

$$s.t. \quad \frac{|x_{k,k}|^2}{\sum_{l \in \mathcal{K}, l \neq k} |x_{k,l}|^2 + \sigma_0^2} \geq \gamma_k^0, \ \forall k \in \mathcal{K} \tag{23b}$$

$$\left( \mathbf{h}_{d,k}^{H} + \mathbf{h}_{r,k}^{H} \mathbf{\Theta} \mathbf{G} \right) \mathbf{p}_j = x_{k,j}, \ \forall k, j \in \mathcal{K} \tag{23c}$$

$$0 \leq \theta_n \leq 2\pi, \ \forall n \in \mathcal{N} \tag{23d}$$

To associate the objective function with the variables, we integrate equality constraints into the objective function of (23) by introducing a penalty parameter [38], yielding the following

optimization problem

$$\mathcal{P'}_{2-\mathrm{II}}: \min_{\{\theta_n\}, \{x_{k,l}\}} \quad \sum_{k \in \mathcal{K}} \|\mathbf{p}_k\|_2^2 + \frac{1}{2\rho} \left( \sum_{k \in \mathcal{K}} \sum_{j \in \mathcal{K}} \left| \left( \mathbf{h}_{d,k}^H + \mathbf{h}_{r,k}^H \boldsymbol{\Theta} \mathbf{G} \right) \mathbf{p}_j - x_{k,j} \right|^2 \right) \tag{24a}$$

$$s.t. \quad \frac{|x_{k,k}|^2}{\sum_{l \in \mathcal{K}, l \neq k} |x_{k,l}|^2 + \sigma_0^2} \geq \gamma_k^0, \ \forall k \in \mathcal{K} \tag{24b}$$

$$0 \leq \theta_n \leq 2\pi, \ \forall n \in \mathcal{N} \tag{24c}$$

In (24), $\rho > 0$ denotes the penalty coefficient used for penalizing the violation of equality constraints. It is worth pointing out that although the equality constraints are relaxed in $\mathcal{P'}_{2-\mathrm{II}}$, when $\rho \to 0 \, (1/\rho \to \infty)$, the solution obtained by solving $\mathcal{P'}_{2-\mathrm{II}}$ always satisfies all equality constraints in $\mathcal{P}_{2-\mathrm{II}}$.

### B. Block Coordinate Descent Algorithm

It can be found in (24) that, for given $\rho > 0$, $\mathcal{P'}_{2-\mathrm{II}}$ is still a non-convex optimization problem due to the non-convex objective function as well as the non-convex constraints. However, it is observed that each optimization variable in $\mathcal{P'}_{2-\mathrm{II}}$ is involved in at most one constraint. This thus motivates us to apply the block coordinate descent method to solve $\mathcal{P'}_{2-\mathrm{II}}$ efficiently by properly partitioning the optimization variables into a pair of blocks.

*1) Subproblem with respect to $\{x_{k,j}\}$:* For any given phase shifts $\{\theta_n\}$, the auxiliary variables can be optimized by solving the following problem (ignoring constant terms)

$$\begin{aligned} \min_{\{x_{k,j}\}} \quad & \sum_{k \in \mathcal{K}} \sum_{j \in \mathcal{K}} |\bar{x}_{k,j} - x_{k,j}|^2 \\ s.t. \quad & \frac{|x_{k,k}|^2}{\sum_{l \in \mathcal{K}, l \neq k} |x_{k,l}|^2 + \sigma_0^2} \geq \gamma_k^0, \ \forall k \in \mathcal{K} \end{aligned} \tag{25}$$

where $\bar{x}_{k,j} = \left( \mathbf{h}_{d,k}^H + \mathbf{h}_{r,k}^H \boldsymbol{\Theta} \mathbf{G} \right) \mathbf{p}_j$ is denoted. In (25), the optimization variables with respect to different receiving users are separable in both objective function and constraints, we can solve the resultant problem by solving $K$ independent subproblems in parallel, each with only one single constraint (i.e., $\frac{|x_{k,k}|^2}{\sum_{l \in \mathcal{K}, l \neq k} |x_{k,l}|^2 + \sigma_0^2} \geq \gamma_k^0$ with $k$ fixed). Specifically, for user $k$, (25) is reduced

to (ignoring constant terms)

$$\min_{\{x_{k,j}\}} \quad \sum_{j \in \mathcal{K}} |\bar{x}_{k,j} - x_{k,j}|^2$$
$$s.t. \quad \frac{|x_{k,k}|^2}{\sum_{l \in \mathcal{K}, l \neq k} |x_{k,l}|^2 + \sigma_0^2} \geq \gamma_k^0. \tag{26}$$

The problem above is a non-convex quadratically constrained quadratic program (QCQP). It has been shown in the literature that strong duality holds for this problem [47]. The Lagrangian function can be expressed as (ignoring constant terms)

$$\mathcal{L}\left(\{x_{k,j}\}, \lambda_k\right) = (1 - \lambda_k) |x_{k,k}|^2$$
$$+ \sum_{l \in \mathcal{K}, l \neq k} \left(1 + \lambda_k \gamma_k^0\right) |x_{k,l}|^2$$
$$- 2 \sum_{j \in \mathcal{K}} \text{Re}\left(\bar{x}_{k,j} x_{k,j}^*\right). \tag{27}$$

Accordingly, the dual function is given by

$$g\left(\lambda_k\right) = \min_{\{x_{k,j}\}} \mathcal{L}\left(\{x_{k,j}\}, \lambda_k\right). \tag{28}$$

It is easy to find that $0 < \lambda_k < 1$; otherwise we have $g\left(\lambda_k\right) \to -\infty$ by setting $|x_{k,k}| \to \infty$. By exploiting the first-order optimality condition, the optimal solution to minimize the Lagrangian function for fixed $\lambda_k$ is given by

$$x_{k,k}^{(*)} = \frac{\bar{x}_{k,k}}{1 - \lambda_k} \tag{29}$$

$$x_{k,l}^{(*)} = \frac{\bar{x}_{k,l}}{1 + \lambda_k \gamma_k^0}, l \in \mathcal{K}, l \neq k. \tag{30}$$

Substituting above optimal solutions into the equality constraint in (26), we obtain

$$f\left(\lambda_k\right) \triangleq \frac{|\bar{x}_{k,k}|^2}{\left(1 - \lambda_k\right)^2} - \sum_{l \in \mathcal{K}, l \neq k} \frac{\gamma_k^0 |\bar{x}_{k,l}|^2}{\left(1 + \lambda_k \gamma_k^0\right)^2} - \gamma_k^0 \sigma_0^2 = 0. \tag{31}$$

It is not difficult to find that $f\left(\lambda_k\right)$ is a monotonically increasing function of $\lambda_k$ for $0 < \lambda_k < 1$. As such, the optimal dual variable and primal variables can be obtained by using the bisection search.

*2) Subproblem with respect to $\{\theta_n\}$:* for any given auxiliary variables $\{x_{k,l}\}$, the phase shifts $\{\theta_n\}$ can be optimized by solving the following problem (ignoring constant terms)

$$\mathcal{P}''_{2-\mathrm{II}} : \min_{\{\theta_n\}} \sum_{k \in \mathcal{K}} \sum_{j \in \mathcal{K}} \left| \mathbf{h}_{r,k}^H \boldsymbol{\Theta} \mathbf{G} \mathbf{p}_j - x_{k,j} \right|^2 \tag{32}$$
$$s.t. \quad 0 \le \theta_n < 2\pi, \ \forall n \in \mathcal{N}.$$

By letting $\varphi_n = e^{j\theta_n}$, $\mathbf{x} = [\varphi_1, \cdots, \varphi_n, \cdots, \varphi_N]^T$, and $\mathbf{g}_j = \mathbf{G} \mathbf{p}_j$, above program is equivalent to

$$\mathcal{P}''_{2-\mathrm{II}} : \min_{\mathbf{x}} \sum_{k \in \mathcal{K}} \sum_{j \in \mathcal{K}} \left| \mathbf{h}_{r,k}^H \mathrm{diag}\left(\mathbf{x}\right) \mathbf{g}_j - x_{k,j} \right|^2 \tag{33}$$
$$s.t. \quad |\varphi_n| = 1, \ \forall n \in \mathcal{N}.$$

In (33), the unit modulus constraints are still included, which are intrinsically non-convex, and there is no general approach to solve the optimization problem with these constraints optimally to the best of our knowledge.

### C. Conjugate Gradient Algorithm Based on Manifold Optimization

In the following subsection, we introduce a conjugate gradient algorithm based on manifold optimization to find a sub-optimal solution to problem (33). Manifold optimization provides a powerful alternative to constrained optimization problem [43]. In literature, the methodology of manifold optimization has been proved to be effective in handling non-convex unit-modulus constraints [32], [44].

To solve (33), we start with some related definitions and terminologies in manifold optimization. A manifold $\mathcal{M}$ is a topological space that resembles a Euclidean space near individual point. The tangent space $T_x \mathcal{M}$ at a given point $x$ on the manifold $\mathcal{M}$ is composed of the tangent vectors $\xi_x$ of the curves $\gamma$ through the point $x$. In most applications, manifolds fall into a special category of topological manifold, namely, a *Riemannian manifold*. A Riemannian manifold is equipped with an inner product, which is defined on the tangent spaces $T_x \mathcal{M}$, and allows one to measure distances and angles on manifolds. More importantly, optimization over a Riemannian manifold is locally analogous to that over an Euclidean space with smooth constraints.

Specifically, the complex circle manifold of $x \in \mathbb{C}$, which is defined by

$$\mathcal{M}_{cc} = \{x \in \mathbb{C} \,|\, x^* x = 1\}. \tag{34}$$

The complex circle manifold $\mathcal{M}_{cc}$ is a Riemannian submanifold of $\mathbb{C}$. Note that the Euclidean metric over the complex plane $\mathbb{C}$ for $\forall x_1, x_2 \in \mathbb{C}$ is defined as $\langle x_1, x_2 \rangle = \mathrm{Re}\left\{x_1^* x_2\right\}$. Hence, the tangent space at the point $x \in \mathcal{M}_{cc}$ can be represented by

$$T_x \mathcal{M}_{cc} = \left\{z \in \mathbb{C} \,|\, \langle x, z \rangle = 0\right\}. \tag{35}$$

Now we expand from a one-dimensional manifold to a multiple dimensional one. Let $\varphi_n = e^{j\theta_n}$ and $\mathbf{x} = [\varphi_1, \cdots, \varphi_n, \cdots, \varphi_N]^T$. Observing the unit modulus constraint $|\varphi_n| = 1$, we may find that $\mathbf{x}$ forms an $N$-dimensional complex circle manifold $\mathcal{M}_{cc}^N = \left\{\mathbf{x} \in \mathbb{C}^N \,|\, x_1^* x_1 = \cdots = x_N^* x_N = 1\right\}$. Intrinsically, the complex circle manifold $\mathcal{M}_{cc}^N$ is a Riemannian submanifold of the $N$-dimensional complex space $\mathbb{C}^N$. Therefore, the feasible region of the optimization problem (33) is over the manifold $\mathcal{M}_{cc}^N$. Further, the tangent space at the point $\mathbf{x} \in \mathcal{M}_{cc}^N$ is expressed as

$$T_\mathbf{x} \mathcal{M}_{cc}^N = \left\{\mathbf{z} \in \mathbb{C}^N \,|\, \mathrm{Re}\left\{\mathbf{z} \circ \mathbf{x}^*\right\} = \mathbf{0}_N\right\}. \tag{36}$$

In the following we introduce some basic operations in manifold optimization.

***Riemannian gradient***: Among all the tangent vectors in $T_\mathbf{x} \mathcal{M}_{cc}^N$, similar to the Euclidean space, one of them is related to the negative Riemannian gradient, representing the direction of the greatest decrease of a function at the point $\mathbf{x} \in \mathcal{M}_{cc}^N$. It can be computed as the projection from the Euclidean gradient $\nabla f(\mathbf{x})$ to the tangent space using the orthogonal projector. Explicitly, the Riemannian gradient at point $\mathbf{x}$ is a tangent vector $\mathrm{grad} f(\mathbf{x})$, which is given by the orthogonal projection of the Euclidean gradient $\nabla f(\mathbf{x})$ onto the tangent space $T_\mathbf{x} \mathcal{M}_{cc}^N$ at point $\mathbf{x} \in \mathcal{M}_{cc}^N$, and can be written as

$$\begin{aligned} \mathrm{grad} f(\mathbf{x}) &= \mathrm{Proj}_\mathbf{x} \nabla f(\mathbf{x}) \\ &= \nabla f(\mathbf{x}) - \mathrm{Re}\left\{\nabla f(\mathbf{x}) \circ \mathbf{x}^*\right\} \circ \mathbf{x}. \end{aligned} \tag{37}$$

Denote the cost function in (33) as $f(\mathbf{x})$, which can be written equivalently as

$$f(\mathbf{x}) = \sum_{k \in \mathcal{K}} \sum_{j \in \mathcal{K}} \left| \sum_{n \in \mathcal{N}} \left(\mathbf{h}_{r,k}^H\right)_n \varphi_n (\mathbf{g}_j)_n - x_{k,j} \right|^2. \tag{38}$$

Specifically, the Euclidean gradient of $f(\mathbf{x})$ is given by

$$\nabla f(\mathbf{x}) = \left[\frac{\partial f(\mathbf{x})}{\partial \varphi_1}, \cdots, \frac{\partial f(\mathbf{x})}{\partial \varphi_n}, \cdots, \frac{\partial f(\mathbf{x})}{\partial \varphi_N}\right]^T$$

where

$$\frac{\partial f(\mathbf{x})}{\partial \varphi_n} = \sum_{k \in \mathcal{K}} \sum_{j \in \mathcal{K}} \left(\mathbf{h}_{r,k}^H \mathrm{diag}(\mathbf{x}) \mathbf{g}_j - x_{k,j}\right)^* \left(\mathbf{h}_{r,k}^H\right)_n (\mathbf{g}_j)_n. \tag{39}$$

Solving the Euclidean gradient involves some techniques on complex-valued matrix derivatives. Interested readers can refer to the details in [45].

**Retraction**: Retraction is another key factor in manifold optimization, which maps a vector from the tangent space back into the manifold. The retraction of a tangent vector $\alpha\mathbf{d}$ at point $\mathbf{x} \in \mathcal{M}_{cc}^m$ is expressed as

$$\text{Rtrctn}_{\mathbf{x}}\left(\alpha\mathbf{d}\right) = \text{vec}\left[\frac{(\mathbf{x} + \alpha\mathbf{d})_i}{|(\mathbf{x} + \alpha\mathbf{d})_i|}\right]. \tag{40}$$

**Transport**: A mapping between two tangent vectors in different tangent spaces is called transport. The transport of a tangent vector $\mathbf{d}$ from $\mathbf{x}_k$ to $\mathbf{x}_{k+1}$ can be expressed as

$$\text{Trnsprt}_{\mathbf{x}_k \to \mathbf{x}_{k+1}}\left(\mathbf{d}\right) = \mathbf{d} - \text{Re}\left\{\mathbf{d} \circ \mathbf{x}_{k+1}^*\right\} \circ \mathbf{x}_{k+1}. \tag{41}$$

---

**Algorithm 1:** Conjugate Gradient Algorithm for Passive Beamforming based on Manifold Optimization

---

    **Input**: $\mathbf{h}_{r,k}, \mathbf{g}_j, x_{k,j}$

    **Output**: $\mathbf{x}$

**1** Randomly initialize $\mathbf{x}_0 \in \mathcal{M}_{cc}^m$; Calculate $\mathbf{d}_0 = -\text{grad}f\left(\mathbf{x}_0\right)$; Set $k = 0$.

**2** **repeat**

**3**      Choose line search step size $\alpha_k$ according to **Goldstein criterion**.

**4**      Find the next point $\mathbf{x}_{k+1}$ using retraction by (40): $\mathbf{x}_{k+1} = \text{Rtrctn}_{\mathbf{x}_k}\left(\alpha_k\mathbf{d}_k\right)$.

**5**      Determine Riemannian gradient at $\mathbf{x}_{k+1}$ by (37) and (38): $\mathbf{g}_{k+1} = \text{grad}f\left(\mathbf{x}_{k+1}\right)$.

**6**      Calculate vector transports $\bar{\mathbf{g}}_k$ of gradient $\mathbf{g}_k$, and $\bar{\mathbf{d}}_k$ of conjugate direction $\mathbf{d}_k$ from $\mathbf{x}_k$ to $\mathbf{x}_{k+1}$ by (41): $\bar{\mathbf{g}}_k = \text{Trnsprt}_{\mathbf{x}_k \to \mathbf{x}_{k+1}}\left(\mathbf{g}_k\right)$, $\bar{\mathbf{d}}_k = \text{Trnsprt}_{\mathbf{x}_k \to \mathbf{x}_{k+1}}\left(\mathbf{d}_k\right)$

**7**      Calculate $\beta_k$ according to **Flecther-Reeves equation**: $\beta_k = (\mathbf{g}_{k+1}^H\mathbf{g}_{k+1})/(\bar{\mathbf{g}}_k^H\bar{\mathbf{g}}_k)$.

**8**      Compute conjugate direction $\mathbf{d}_{k+1} = -\mathbf{g}_{k+1} + \beta_k\bar{\mathbf{d}}_k$.

**9**      Update $k \leftarrow k + 1$.

**10** **until** *convergence*;

**11** Return $\mathbf{x} = \mathbf{x}_{k+1}$.

---

Bearing the Riemannian gradient, retraction and transport in mind, we are able to develop a conjugate gradient method in Riemannian space. The procedures of obtaining the passive beamformer $\{\theta_n\}$ based on manifold optimization as well as conjugate gradient method are listed in Algorithm 1. Algorithm 1 utilizes the well-known Goldstein criterion and Flecther-Reeves equation to guarantee the objective function to be non-increasing in each iteration. According to Theorem 4.3.1 in [46], Algorithm 1 is guaranteed to converge to a critical point, i.e., at this point the gradient of the objective function is zero.

---

**Algorithm 2:** Alternating Optimization Algorithm

---

**Input**: $\{\mathbf{h}_{d,k}\}$, $\{\mathbf{h}_{r,k}\}$, $\mathbf{G}$ and $\{R_k^0\}$

**Output**: $\{c_f\}$, $\{\mathbf{p}_k\}$ and $\{\theta_n\}$

1 Randomly construct the initial point $\left\{\mathbf{p}_k^{(0)}\right\}$ and $\left\{\theta_k^{(0)}\right\}$; Set the convergence tolerance $\epsilon > 0$ and iteration index $t = 0$ for the outer layer.

2 Find $\{c_f\}$ by solving problem (5).

3 **repeat**

4 $\quad$ Solve (9) to obtain $\mathbf{p}_k^{(t+1)}$ for $\theta_n = \theta_n^{(t)}$.

5 $\quad$ **repeat**

6 $\quad\quad$ Update the auxiliary variables $\{x_{k,j}\}$ by solving (25).

7 $\quad\quad$ Update the complex circle manifold vector $\mathbf{x}$ by running Algorithm 1.

8 $\quad$ **until** *The decrease of the objective value of* (24) *is below a threshold* $\delta > 0$;

9 $\quad$ $\theta_n^{(t+1)} = -j \ln \{\mathbf{x}\}_n$.

10 $\quad$ Update $t \leftarrow t + 1$.

11 **until** *The decrease of the objective value of* (4) *is below the threshold* $\epsilon$;

12 Return $\{c_f\}$, $\left\{\mathbf{p}_k = \mathbf{p}_k^{(t+1)}\right\}$ and $\left\{\theta_n = \theta_n^{(t+1)}\right\}$.

---

### D. Overall Algorithm and Complexity Analysis

Based on the above analysis, the overall alternating optimization algorithm proposed in this paper is summarized in Algorithm 2. At first, the content placement problem is solved in Step 2. Step 3 to Step 11 are performed to optimize active and passive beamforming alternately. Step 5 to Step 8 invoke the block coordinate descent method to deal with the passive beamforming integrally, where in each iteration, Step 6 exploits bisection search to solve the auxiliary variables, and Step 7 resorts to conjugate gradient algorithm to compute present phase shifts.

Let us analyze the complexity of the overall algorithm. The SDP problem (9) can be solved by iterative optimization techniques, e.g., interior-point method (IPM). A worst-case complexity result to solve the SDP is given by [48]

$$\mathcal{O}\left(\sqrt{\sum_{i=1}^{N_{sdp}} n_i^{sdp} + m}\left(\sum_{i=1}^{N_{sdp}} \left(n_i^{sdp}\right)^3 + \sum_{i=1}^{N_{sdp}} \left(n_i^{sdp}\right)^2 m + m^3\right) \log\left(1/\varepsilon_1\right)\right) \tag{42}$$

where $N_{sdp}$ is the number of SDP cone constraints, $n_i^{sdp}$ is the dimension of the $i$-th SDP cone, $m$ is the number of constraints, and $\varepsilon_1$ is the accuracy of the convex optimization solution. With regards to the SDP considered in this paper, its worst-cast complexity is correspondingly given by $\mathcal{O}\left(\sqrt{KM + K}\left(KM^3 + K^2M^2 + K^3\right)\log\left(1/\varepsilon_1\right)\right)$, i.e., the complexity of Step 4 in

Algorithm 2. For the Step 6 in Algorithm 2, it can be shown that the complexity of solving (25) is $\mathcal{O}\left(K^2 \log\left(1/\varepsilon_2\right)\right)$, where $\varepsilon_2$ is the accuracy for the bisection search. For the conjugate gradient algorithm based on manifold optimization, i.e., Algorithm 1, for each iteration, the complexity mainly depends on the calculation of the Riemannian gradient, which is given by $\mathcal{O}\left(N^2\right)$. Denote the iteration number required by the conjugate gradient algorithm as $T$. Then the complexity of Step 7 in Algorithm 2 is given by $\mathcal{O}\left(TN^2\right)$. Thus, the overall complexity of Algorithm 2 can be written as

$$\mathcal{O}\left(I_{out}\sqrt{KM+K}\left(KM^3+K^2M^2+K^3\right)\log\left(1/\varepsilon_1\right)I_{in}\left(K^2\log\left(1/\varepsilon_2\right)+TN^2\right)\right)$$

where $I_{in}$ and $I_{out}$ denote the outer and inner iteration numbers of Algorithm 2 required for convergence, respectively.

## V. SIMULATION ANALYSIS

### A. Simulation Settings

We consider a system operating on a carrier frequency of $2.4$ GHz. The noise power spectral density is given by $\sigma_0^2 = -150$ dBm/Hz, and bandwidth is $B = 10$ MHz. As shown in Fig. 2, a three-dimensional (3D) coordinate system is considered, where the BS is equipped with a uniform linear array (ULA) located on the $x$-axis, and the RIS is configured with a uniform planar array (UPA) located on the $y - z$ plane, respectively. The reference locations of the BS and the RIS are set at $(5, 0, 30)$ meters and $(0, 5, 10)$ meters, respectively. The antenna spacing is half wavelength. The served users $k \in \mathcal{K}$ are randomly and uniformly distributed in a circle region centered at $(5, 10, 1.5)$ with a radius of $2.5$ m. Here the $z$-coordinates indicate that the heights of BS, RIS, and users are assumed to be $30$ m, $10$ m, and $1.5$ m, respectively. The number of transmit antennas at the BS is $M = 16$, and the number of serving users is $K = 5$. We consider $F = 1000$ files in the database, and $S_0 = 100$ for the local storage size at the BS. Unless otherwise stated, the number of the reflection elements is $N = 50$, and Zipf parameter is set to $\varepsilon = 1$.

The distances for the direct BS-user link, the BS-RIS link and the RIS-user link are denoted by $d_{d,k}$, $d_G$ and $d_{r,k}$, respectively. The distance-dependent path loss for all channels is modeled as $PL(d) = \rho_0\left(\frac{d}{d_0}\right)^{-\alpha}$, where $\rho_0 = -30$ dB denotes the path loss at the reference distance $d_0 = 1$ m, $d$ denotes the link distance [32], and $\alpha$ denotes the path loss exponent. For small

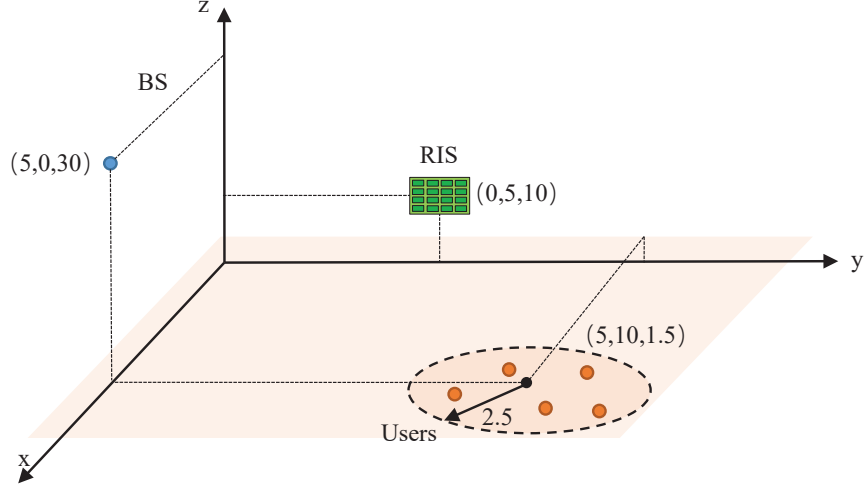

Fig. 2. Simulation scenario.

scale fading, the Rayleigh fading and the Rician fading models are assumed for the direct BS-user link and the BS-RIS/RIS-user links, respectively [34]. Then, the corresponding channel coefficients can be expressed as

$$\mathbf{h}_{d,k} = \sqrt{PL(d_{d,k})} \mathbf{h}_{d,k}^{NLOS} \tag{43}$$

$$\mathbf{G} = \sqrt{\frac{PL(d_G)}{K_G+1}} \left( \sqrt{K_G} \mathbf{G}^{LOS} + \mathbf{G}^{NLOS} \right) \tag{44}$$

$$\mathbf{h}_{r,k} = \sqrt{\frac{PL(d_{r,k})}{K_{r,k}+1}} \left( \sqrt{K_{r,k}} \mathbf{h}_{r,k}^{LOS} + \mathbf{h}_{r,k}^{NLOS} \right) \tag{45}$$

where $K_G$ and $K_{r,k}$ denote the related Rician factors, $\mathbf{G}^{LOS}$ and $\mathbf{h}_{r,k}^{LOS}$ denote the deterministic line-of-sight (LoS) components, $\mathbf{h}_{d,k}^{NLOS}$, $\mathbf{G}^{NLOS}$ and $\mathbf{h}_{r,k}^{NLOS}$ denote the non-LoS fading components. In specific, the non-LoS component is modeled as Rayleigh fading, while the LoS component is modeled as the product of the array response vectors of the transceivers [32], [34]. For instance, $\mathbf{G}^{LOS}$ is given by

$$\mathbf{G}^{LOS} = \mathbf{a}_{\mathrm{G}}\left(\theta\right) \mathbf{a}_{\mathrm{BS}}^{H}\left(\phi\right)$$

with

$$\mathbf{a}_{\mathrm{G}}\left(\theta\right) = \left[ 1, e^{j\frac{2\pi}{\lambda}d\sin(\theta)}, \cdots, e^{j\frac{2\pi}{\lambda}d(N-1)\sin(\theta)} \right]^{T}$$

$$\mathbf{a}_{\mathrm{BS}}\left(\phi\right) = \left[ 1, e^{j\frac{2\pi}{\lambda}d\sin(\phi)}, \cdots, e^{j\frac{2\pi}{\lambda}d(M-1)\sin(\phi)} \right]^{T}$$

where $\theta$ and $\phi$ are the angles of arrival and departure (AoA/AoD), respectively, $\lambda$ is the signal wavelength, and $d$ is the distance between antenna elements. In this paper, without otherwise specified, the path loss exponents for the direct link, BS-RIS and RIS-user link are set to be $\alpha_{d,k} = 3.5$, $\alpha_G = 2.2$ and $\alpha_{r,k} = 2.2$, respectively. The target SINR requirements for different users are all set to $\gamma_0 = 30$ dB, which is related to the content-delivery rate of $100$ Mbps. All simulation results are obtained by averaging over 1000 independent realizations.

The following heuristic caching strategies are considered in simulations.

- *Uniform random caching (URC)*: In each realization, BS caches the content files randomly with equal probabilities regardless of their popularity distribution.

- *File popularity based probabilistic caching (FPPC)*: In each realization, BS caches a content randomly with probability depending on the content popularity, and the more popular the content is, the more likely it will be cached.

- *Optimized caching (OC)*: In each realization, the content placement is optimized by solving the program $\mathcal{P}_1$.

The following three communication schemes are considered for comparisons.

- *With RIS*: The active beamforming at BS and the phase shifts of the reflecting elements on RIS are optimized by invoking Algorithms 1 & 2.

- *Random phase*: The phase shifts of the reflecting elements on RIS are randomly generated, whilst the active beamforming is designed by solving the program $\mathcal{P}_{2-I}$.

- *Without RIS*: The reflecting path $\mathbf{h}_{r,k}^H \boldsymbol{\Theta} \mathbf{G}$ is set to zero. The active beamforming is designed by solving the program $\mathcal{P}_{2-I}$.

### B. Simulation Results

Figure 3 compares the power cost performance of three communication schemes versus the number of reflecting elements $N$, where the OC strategy is applied in these schemes. Specifically, it is seen that the scheme without RIS endures the highest power cost, indicating that applying RISs can effectively decrease the power consumption. For the pair of schemes with RIS, the power cost can be reduced by increasing $N$. Notice that installing more passive reflecting elements is practical, and both energy and cost efficient, since RISs do not need radio frequency chains, and are compatible with the hardware of existing wireless networks. Besides, it is observed that the performance of random phase is inferior to that of the RIS scheme with optimized phase shifts. This is because the reflected signals have not been constructively added towards the target receivers. By contrast, with aid of our proposed passive beamforming design, the power cost has been decreased dramatically.

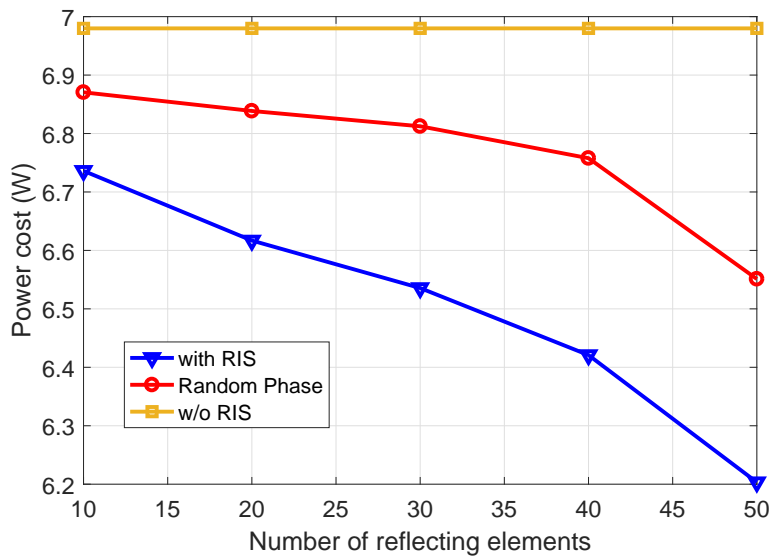

Fig. 3. Power cost versus the number of reflecting elements.

Figure 4 compares the backhaul capacity of caching strategies versus the Zipf exponent $\varepsilon$. Clearly, compared with the benchmark of no caching, all caching schemes achieve much lower backhaul capacity cost. Notice that with the increase of the Zipf exponent, both of the content popularity and the user requests are concentrated on fewer files. It is observable from Fig. 4 that, upon increasing Zipf exponent, the backhaul capacity of URC strategy stays the same value. This is because in URC strategy, BS caches content files randomly with equal probabilities regardless of the content popularity or the request distribution. By contrast, the backhaul capacities of FPPC and OC keep shrinking with the increase of the Zipf exponent. Notably, our proposed OC strategy achieves the lowest backhaul cost all the time, demonstrating that our proposed content placement design is capable of reaping the benefit of skewed content popularity and user request distribution.

Denote the coordinate in y-axis of RIS as $y_{\mathrm{RIS}}$. In Fig. 5, we study the impact of the RIS location by moving the RIS from $y_{\mathrm{RIS}} = 0$ m to $y_{\mathrm{RIS}} = 10$ m. It is observed that the power consumed by the RIS scheme first decreases with $y_{\mathrm{RIS}}$, and then increases for $y_{\mathrm{RIS}} > 5$ m. This is because the large scale path loss of BS-RIS-user links is $PL(d_G)PL(d_{r,k})$, which achieves its minimum around the middle of the y-coordinate and can be calculated accurately in the 3D coordinate system. As RIS is very close to the destination in y-axis, the power cost decreases slightly, since the RIS-user link becomes much better. By contrast, moving the RIS location along with the y-axis fails to bring an obvious variation trend to the random phase scheme. This is because, with randomly generated phase shifts, the RIS cannot reflect signals to beam effectively towards the destination, not to mention adapting to the RIS location. Nevertheless, both RIS schemes achieve better power cost performance than the no RIS one.

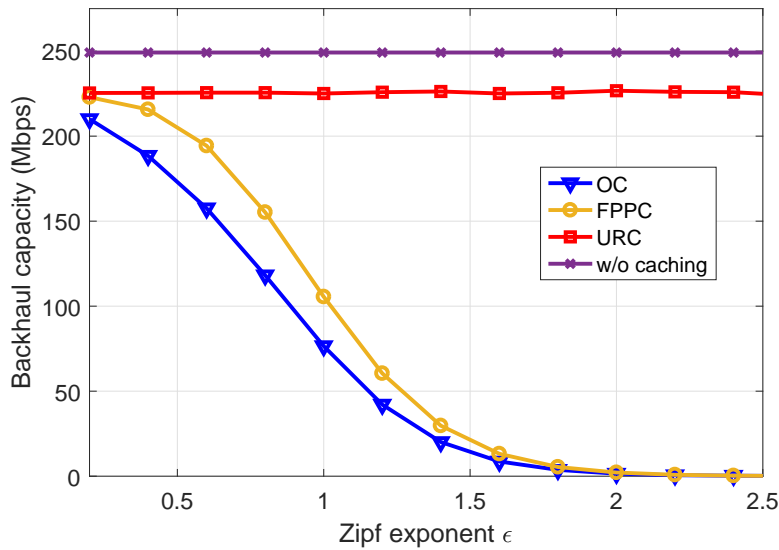

Fig. 4. Backhaul capacity versus Zipf exponent.

In previous simulations, the path loss exponents of the RIS-related links are set to $\alpha_G = 2.2$ and $\alpha_{r,k} = 2.2$, which are close to the propagation conditions in free space. However, it may become impractical in some specific scenarios with certain obstacles. Hence, we are motivated to investigate the impact of the RIS-related path loss exponents on the achievable performance. In Figs. 6 and 7, the impacts of the path loss exponents of BS-RIS and RIS-user links are depicted, respectively. It is observed that the power cost achieved by the proposed scheme increases upon increasing $\alpha_G$ and $\alpha_{r,k}$, and finally approaches the power cost as achieved by the no RIS scheme. This is because, upon increasing $\alpha_G$ and $\alpha_{r,k}$, the signal attenuation associated with the RIS-related links becomes larger, leading to more power consumption at the transmitter. However, when $\alpha_G$ and $\alpha_{r,k}$ are small, a significant performance gain can be achieved by our proposed scheme compared with random phase and no RIS schemes. This observation reminds us that an RIS had better to be deployed in a relatively open scenario with few obstacles so as to reap more performance gains.

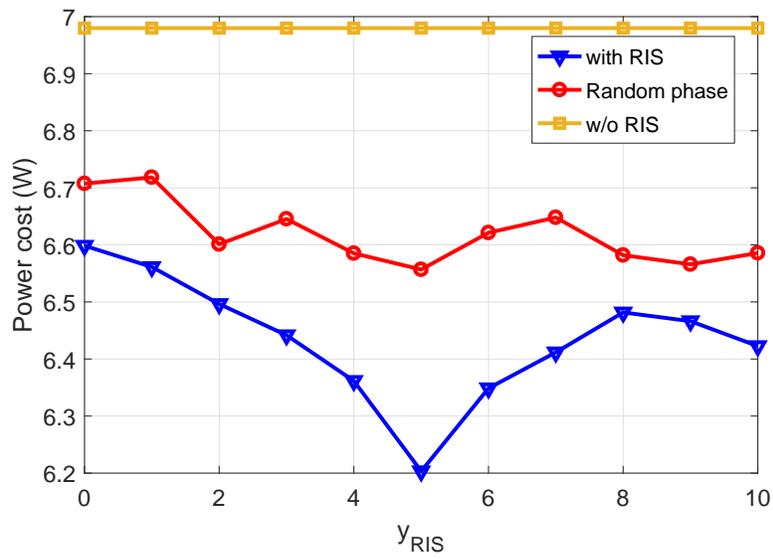

Fig. 5. Power cost versus the RIS location.

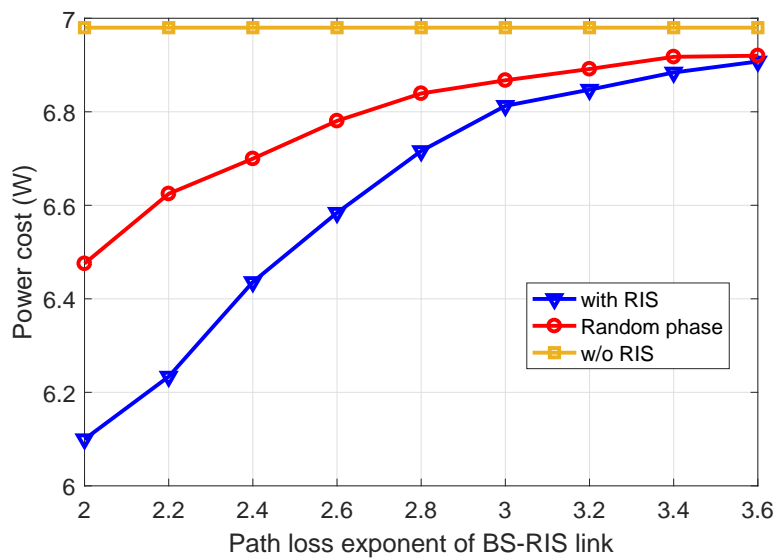

Fig. 6. Power cost versus the path loss exponent of BS-RIS link.

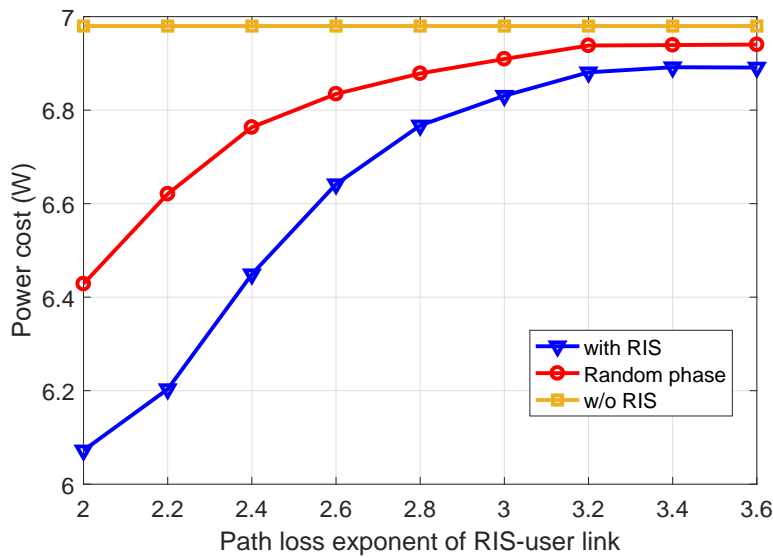

Fig. 7. Power cost versus the path loss exponent of RIS-user link.

## VI. CONCLUSIONS

In this paper, an RIS-aided edge caching system has been considered, where a network cost minimization problem has been formulated to optimize content placement and hybrid beamforming. After decoupling the content placement subproblem with the hybrid beamforming design, we have proposed an alternating optimization algorithm to tackle the active beamforming and passive phase shifting. For active beamforming, we have transformed the problem into an SDP by applying SDR. For passive phase shifting, we have introduced block coordinate descent method to alternately optimize the auxiliary variables and the RIS phase shifts. Further, a conjugate gradient algorithm based on manifold optimization has been proposed to deal with the non-convex unit-modulus constraints in the passive phase shifting design. Numerical results have showed that our RIS-aided edge caching design can effectively decrease the network cost in terms of backhaul capacity and power consumption, compared with existing caching strategies, random phase shifting scheme, and no RIS counterpart.